\documentclass{svproc}
\usepackage{url}
\usepackage[utf8]{inputenc}
\usepackage{mathptmx}       % selects Times Roman as basic font
\usepackage{helvet}         % selects Helvetica as sans-serif font
\usepackage{courier}        % selects Courier as typewriter font
\usepackage{type1cm}        % activate if the above 3 fonts are
\usepackage{makeidx}         % allows index generation
\usepackage{graphicx}        % standard LaTeX graphics tool
\usepackage{multicol}        % used for the two-column index

\usepackage[bottom]{footmisc}% places footnotes at page bottom%% Useful packages

\usepackage{amsmath}
\usepackage{multirow}
\usepackage{graphicx}
\usepackage{caption}
\usepackage{subcaption}
\begin{document}
\mainmatter              % start of a contribution
\title{Systematic Biases in Link Prediction: comparing heuristic and graph embedding based methods}
\titlerunning{Systematic Biases in Link Prediction}  % abbreviated title (for running head)
\author{Aakash Sinha\inst{1} \and Rémy Cazabet \inst{2}\and Rémi Vaudaine\inst{3}}
\authorrunning{Aakash Sinha et al.} % abbreviated author list (for running head)
%Craig Chambers, Kim B. Bruce, and Elisa Bertino}
%
\institute{Department of Computer Science and Engineering, I.I.T. Delhi, Hauz Khas, New Delhi - 110016, INDIA
\and
Univ Lyon, UCBL, CNRS, LIRIS UMR 5205, F-69621, Lyon, France
\email{remy.cazabet@gmail.com}
\and
Univ Lyon, UJM-Saint-Etienne, CNRS, SAINT-ETIENNE, France
}

\maketitle              % typeset the title of the contribution

\begin{abstract}
Link prediction is a popular research topic in network analysis. In the last few years, new techniques based on graph embedding have emerged as a powerful alternative to heuristics. In this article, we study the problem of systematic biases in the prediction, and show that some methods based on graph embedding offer less biased results than those based on heuristics, despite reaching lower scores according to usual quality scores. We discuss the relevance of this finding in the context of the \textit{filter bubble} problem and the \textit{algorithmic fairness} of recommender systems.
\keywords{graph embedding, link prediction, systematic biases, filter bubble}
\end{abstract}

\section{Introduction}

Graph data occurs in various real-world applications such as social networks, biological networks, communication networks and many more. 

In this article, we focus on the problem of \textit{link prediction} (LP). LP is a classic problem on graph data, with numerous applications, from the identification of missing data to recommender systems. Several surveys have been written on the topic \cite{survey_link2,lichtenwalter2010new,survey_link1,LP_eval}. These articles have notably discussed the best way to evaluate and compare LP methods, and have studied the properties, advantages and drawbacks of different approaches. Studied approaches can be \textbf{unsupervised} or \textbf{supervised}, and be based on \textbf{heuristics} (e.g. the number of common neighbors, Adamic Adar index, etc.) or other approaches (block modeling,  random walks, etc.).

Graph embedding is a new technique that gained significant popularity amongst the research community in recent years. Embeddings convert graph data into vectors, creating a representation of nodes in a lower dimensional vector space, on top of which various prediction and detection algorithms can be applied. These embeddings can be used, for instance, in node classification \cite{nodeclassification,Goyalsurvey}, community detection \cite{communitydetection,verse} and role detection \cite{ribeiro2017struc2vec}. Recent surveys \cite{survey2,Goyalsurvey,cui2018survey} present in details the rationale, the different methods and applications of graph embeddings.

In recent years, several articles (e.g.,\cite{verse,node2vec,Goyalsurvey}) have proposed to use \textbf{graph embedding} for link prediction. Such papers often claim that these approaches outperform the state of the art. It must be noted, however, that these articles are focused on proposing new embedding approaches, and do not proceed to in-depth evaluation. In particular, they often do not use the quality scores recommended in the literature, and do not investigate the results behind computing such a score.

This article is organized in three parts
\begin{enumerate}
	\item Using a rigorous evaluation framework, we compare quantitatively recent approaches based on graph embeddings on the task of link prediction with earlier methods
	\item We evaluate systematic biases introduced by each approach according to three aspects: the distance in the graph, the degrees of nodes and the community structure
	\item We discuss the potential effect of the observed biases in the context of the \textit{filter bubble} problem and the \textit{algorithmic fairness} of recommender systems.
\end{enumerate}

%In the first section, we will present an evaluation framework that can be used to compare any link prediction method, independently of its approach. The second section presents the results of our experiments, showing that embedding-based methods do not outperform heuristic-based ones in most settings. Finally, in the last section, we present a way to combine methods based on heuristics and embeddings, and show that it outperforms the state of the art of both of them used separately.

\section{Link Prediction Evaluation Framework}

In this section, we define a rigorous framework to compare link prediction methods. According to our experience, many recent articles on link prediction do not specify precisely their experimental sections, leading to great difficulties in reproducing the reported results. We split the evaluation framework into three independent steps, each described in the corresponding section:
\begin{enumerate}
	\item Creation of training and test sets
	\item Link Prediction
	\item Evaluation using an appropriate score function
\end{enumerate}

\subsection{Creation of learning and prediction sets}

%For most machine learning problems, we create datasets, split it into train and test set. The train data is then used to learn the models and test data is used to evaluate the model. 

%If we already construct the dataset of positive and negative examples, we tend to introduce a bias in a way that we already know the future. Instead what must be done is that we should keep the test graph completely separate, and get the score for all edges not in the train graph. And then compute the performance of the model using the test and the predicted edge scores.

%The evaluation process should go in the following way: split the graph into train and test, and keep aside the test edges. Considering the prediction algorithm as a black box, what it should do is produce scores for each edge not already present in the train graph i.e. n*n-train edges. Now this should be used against the test to compute the performance.

To evaluate link prediction methods, one starts from a network dataset, and split it into a \textbf{learning set} and a \textbf{prediction set}, even if the method to test is considered as unsupervised. The learning set is considered the \textit{current state} of the network, based on which predictions are made. The prediction set is considered as the \textit{future} evolution of the network, i.e. the list of edges that will appear and that should be predicted. In this article as in most of the literature, we focus only on \textit{added} edges, and not \textit{removed} ones.

%The chosen dataset can be of two types: 
%\begin{itemize}
%	\item Static graph
%	\item Dynamic graph
%\end{itemize}

Note that since methods are \textit{supervised}, they need to split the learning set into a training and test set in order to learn where edges should appear. This is independent of the described distinction between learning and prediction set. In real applications, the input is an observed graph --that can be static or dynamic-- and that graph corresponds to our learning set.

\subsubsection{Static graph}
If the original dataset used for evaluation is a single, static graph, we remove randomly a fraction of edges of a given size. These removed edges constitute the \textbf{prediction set}. Following a common practice \cite{Goyalsurvey}, we ensure that the resulting network is composed of a single connected component, using the following procedure:
\begin{itemize}
	\item Remove randomly the desired number of edges
	\item Conserve only the largest component of the resulting graph
\end{itemize}

\subsubsection{Dynamic graph}
For dynamic graphs, the order of edge apparition is known. A date is chosen to split the dataset in two: all edges appearing before or at the chosen date constitute the learning set, and all those appearing after the date constitute the prediction set. If the original data contains information on edges that disappear, this information is ignored. To stay coherent with the static case, the single connected component constraint also applies: in the learning set graph, only the largest connected component is conserved.

\vspace{0.5cm}
In our experiments, the learning set is composed of \textbf{80\%} of the original dataset.

\subsection{Link Prediction}
%The link prediction step is specific to each algorithm. 

%For unsupervised methods, a score is computed for each pair of unconnected nodes in the learning set. The greater the score, the higher the probability to see an edge appear between them. A simple example of score is the \textit{number of common neighbors}. 

A classifier is trained from examples to recognize the pairs of nodes that are the most likely to see edges appear between them. Examples are composed of 50\% of \textit{positive} examples (i.e. examples of pairs of nodes that are connected by a link) and 50\% of \textit{negative} examples (i.e. pairs of nodes that are \textbf{not} connected by a link). Both positive and negative examples are picked randomly from the learning set graph. In this article, we pick $\frac{1}{4}$ of edges in the learning set as positive examples, and an equal number of non-edges as negative examples.

The trained classifier could be used to predict, given an unseen pair of nodes, if it should be connected by an edge or not (Yes/No prediction). However, this does not make sense in realistic settings, since we often want to answer questions such as \textit{which edges might appear in the next hours/day/month ?} or \textit{What are the three most likely edges to appear connecting node $n_x$ ?}. As a consequence, following previous works \cite{survey_link2,node2vec}, we assign to each pair of nodes a score corresponding to the \textit{decision function} of the classifier for this sample.

%We can note that using this approach, unsupervised and supervised methods can be compared and evaluated using a common process since both of them provide a similar lin prediction on the form of an ordered list of probable edges to appear.

\vspace{0.5cm}
For each experiment, we run the link prediction process 5 times and report the average and standard deviation values.

\subsection{Choice of an appropriate score function}
%Different works evaluate their models in very different way which leads to incorrect ranking of methods. In order to produce a correct ranking of methods we need to have a standard way to evaluate link prediction models. In this section, we describe a method to evaluate link prediction algorithms. We also argue against the construction of mutually exclusive train and test datasets...

%To evaluate a model correctly, the prediction should be made on all possible |V|*|V|-|E| edges. Any under sampling will lead to unwanted bias in measurement leading to incorrect ranking. Since this test size can be very large for real graphs, we need to reduce the size of the test set without introducing any or minimal bias.

%What must be kept in mind while sampling the test set is that the ratio of positive to negative edges must be kept the same as is in the original test set. Often researchers tend to under sample negative examples to balance the test set. This kind of evaluation leads to a deviation from the real scenario and hence produces incorrect rankings. 

%In real life, what would happen is that we would only be given a graph and the future edges would be hidden from us. From this graph we need to predict the score for all possible edges. Thus the model evaluation process should also proceed in this way so that we can correctly say evaluate the model. The score which we will get this way would be a representative of the score which we would get if we do this task on a real time evolving network.

To evaluate the quality of the prediction provided by a link prediction method, it is necessary to use a relevant score function. 
%A score function takes as  input:
%\begin{itemize}
%	\item The ordered list of prediction for each pair of nodes not connected in the learning set
%	\item The ground truth label for this pair of nodes, i.e. 1 if this pair of nodes is connected by an edge in the prediction set, 0 otherwise
%\end{itemize} 

It must be noted that link prediction is characterized by an extreme class imbalance. On realistic networks of large size, the \textit{density} is low ($<$0.001\%), thus the number of edges that will appear is much lower than the number of edges that will not appear.

To evaluate a model correctly, the prediction should be made on all possible $|V|*|V|-|E|$ edges. Since this test size can be very large for real graphs, we need to reduce the size of the test set by taking a random sample without introducing any or minimal bias.
%For practical reasons, on large graphs, the evaluation is conducted on a random sample. 
In this article, we fix a sample size of 500 000. Two important observations must be made about the constitution of this sample:
\begin{itemize}
	\item Most scores yield very different results according to the fraction of negative examples in the sample. Unlike for classifier training, it is thus essential\cite{LP_eval} to keep its original ratio, and not take a 50\%/50\% sample, which is highly unrealistic.
	\item Due to the extreme imbalance, the number of positive examples can be extremely low even in a large sample, leading to noise in the results. Therefore we fix a lower value on the number of positive samples (10 in this article).
\end{itemize}

Two metrics, the Average Precision (AP) and the Area Under Receiver Operating Characteristic curve (AUROC) have been identified \cite{LP_eval,survey_link1} to be relevant for this task. AUROC has the property of being independent of the fraction of positive examples in the test set, while AP is favored by some because it gives a higher importance to the first few predictions, that are the most useful in link prediction settings.

\section{Methods evaluation}

In this article, we present results on three graphs previously used in articles on graph-embedding techniques. 

We selected large graphs, the first two being static and the last one dynamic. These graphs are FACEBOOK \cite{datasetFB}, ASTROPH \cite{datasetASTRO} and VK \cite{verse}. Due to space constraint, the reader can refer to their original descriptions in the referenced articles.

%Table \ref{tab:graphs} lists the datasets used for evaluation along with some properties. FACEBOOK has notably been used for evaluation 
%\begin{table}
%\centering
%\begin{tabular}{l|l|l|l}
%Name & |V| & |E| & Density \\\hline
%FACEBOOK \cite{datasetFB} & 4k & 88k & 0.0055 \\
%ASTROPH \cite{datasetASTRO} & 18k & 198k & 0.00061 \\
%VK \cite{verse} & 79k & 2.7M & 0.00043 \\
%%CoCit \cite{datasetCoCit} & 44k & 195k & 0.0001 \\
%\end{tabular}
%\caption
%{\label{tab:graphs}Datasets used for evaluation. VK is a dynamic graph.}
%\end{table}

%\subsection{Methods}
%We evaluate and compare methods based on heuristics and methods based on graph embeddings. For both of them, we use supervised and unsupervised prediction techniques. 
%
\subsection{Method based on Heuristics}
The heuristics used in all experiments are Common Neighbors, Adamic Adar,  Preferential attachment, Jaccard Coefficient, nodes degree (for both endpoints) and Resource allocation index (Table \ref{tab:heuristics}).
%It must be noted that all of them but Preferential Attachment and degrees can give non-zero scores only for pairs of nodes at distance 2 in the graph, i.e. that have at least a common neighbor.

Heuristics can be used for unsupervised prediction if a single of them is used: the score of the heuristic is used directly to rank pairs of nodes. They can also be used for supervised prediction by using some or all of them as features, together with a classifier. In this article, we use all heuristics together and a logistic classifier (implementation of sklearn). This supervised approach has been shown to give the best results \cite{al2006link}. 

\begin{table}
\centering
\begin{tabular}{l|c}
Heuristics & Definition \\\hline
Common Neighbors & $|\Gamma(u) \cap \Gamma(v)|$ \\
Adamic Adar & $\sum\limits_{w \in \Gamma(u) \cap \Gamma(v)} \frac{1}{\log |\Gamma(w)|}$ \\
Preferential attachment & $|\Gamma(u) * \Gamma(v)|$ \\
Jaccard Coefficient & $\frac{|\Gamma(u) \cap \Gamma(v)|}{|\Gamma(u) \cup \Gamma(v)|}$ \\
Resource allocation index  & $\sum\limits_{w \in \Gamma(u) \cap \Gamma(v)} \frac{1}{|\Gamma(w)|}$ \\

\end{tabular}
\caption
{\label{tab:heuristics}Heuristic scores for Link prediction. u,v represent the nodes, $\Gamma(u)$ represents the set of neighbors of node u.}
\end{table}

\subsection{Methods based on Graph embeddings}
%Graph embeddings provide a representation of nodes as vectors. These vectors can be used to compute scores for possible edges in a supervised way or in an unsupervised way. For unsupervised prediction, the score is calculated directly using a distance function specific to each method (Euclidean or cosine). This method has been used, for instance, in \cite{Goyalsurvey}.
%
%For supervised prediction, the vectors of the two nodes are combined using an appropriate operation (see Table \ref{tab:metrics}) to obtain a vector characterizing the edge. In the literature, several operations are tested to combine vectors, and the best one is selected. However, this approach raises concerns about hyper-parameter overfitting, and we therefore use only the one that has been observed to provide the best results for each approach. A classifier is subsequently trained on the edge vectors. We use the same implementation of logistic classifier than for heuristics.
%
%We chose to consider four graph embedding techniques with available implementations, that, according to several authors \cite{Goyalsurvey,verse,node2vec}, provide state-of-the-art results for link prediction.
%
%
%Formally, we define a graph embedding as:
%
%\begin{definition}
%Given a graph G (V,E), an embedding algorithm outputs a function f : v -> x ∈ $R_d$ for all v ∈ V such that d<<|V|. The function f corresponds to the property being preserved by the embedding. 
%\end{definition}

Different embedding algorithms preserve different properties (first/second/higher order proximities) and capture different features. The four embeddings we test are:
\begin{itemize}
	\item Laplacian Eigenmaps (LE) \cite{LE}
	\item High-Order Proximity preserved Embedding (HOPE)\cite{HOPE}
	\item node2vec\cite{node2vec}
	\item Versatile Graph Embeddings from Similarity Measures (VERSE)\cite{verse}
\end{itemize}

Different operators are used to combine the vectors of nodes into edge vectors: \textit{Hadamard} for LE and VERSE, \textit{Normalized Hadamard} for HOPE and node2vec  These operators are those observed by the authors to yield the best results. We use embeddings in 128 dimensions, as is a common practice in the literature.

Due to space constraints, we refer the reader to the original article for each method detailed description. We have chosen these methods due to their frequent use and good results observed in the recent literature, e.g., \cite{Goyalsurvey}. We have used the most common parameters. In particular for node2vec, we have systematically tested with parameter sets (p=4,q=0.5), (p=0.5, q=4), (p=1,q=1) and we report the highest score.

%\paragraph{Laplacian Eigenmaps}
%
%
%Laplacian Eigenmaps aims to keep the distance between two nodes inversely proportional to the weight of the edge. Whenever $W_{uv}$ is high, it keeps the nodes close, thereby preserving first order proximity.
%In particular, LE minimizes the following objective function:

%\begin{table*}[h]
%\centering
%\begin{tabular}{l|l|l|l|l}
%\textbf{Method} & LE & HOPE & node2vec & VERSE \\\hline
%%Unsupervised & Euclidean distance & Cosine similarity & Cosine similarity & Dot product \\
%\textbf{Operator} & Hadamard & Normalised  Hadamard & Normalised  Hadamard & Hadamard
%\end{tabular}
%\caption{\label{tab:metrics}operators used to combine node vectors, for each algorithm.}
%\end{table*}

\subsection{Results}

Each method has been tested on each graph. Results are summarized in Table \ref{tab:results}. We can observe that, compared with state of the art supervised heuristics approaches, graph embeddings do not yield clear better results. Using the AP score, heuristics always perform best. Using the ROC score, the VERSE algorithm is the only algorithm to obtain higher results.

\begin{table}
\centering
\begin{tabular}{l|l|l|l|l|l|l}
Graph & Method &                     Heuristics & LE   & HOPE & n2v & VERSE \\\hline
\multirow{3}{*}{AP} & FACEBOOK & \textbf{0.74} & 0.50 & 0.68 & 0.44 & 0.60 \\
					& ASTROPH & \textbf{0.79} & 0.06 & 0.43 & 0.22 & 0.48 \\
					& VK & \textbf{0.063} & - & - & 0.006 & 0.021 \\
					\hline

\multirow{3}{*}{ROC} & FACEBOOK &\textbf{0.995} & 0.994 & 0.981 & 0.988 & \textbf{0.995}\\					& ASTROPH &0.988 & 0.922 & 0.943 & 0.973 & \textbf{0.992}\\
					&VK &0.87 & - & - & 0.77 & \textbf{0.89}\\

%\multirow{3}{*}{ROC} & AP & \textbf{0.79} & 0.06 & 0.43 & 0.22 & 0.48 \\
%						  & ROC &0.988 & 0.922 & 0.943 & 0.973 & \textbf{0.992}\\
%\multirow{3}{*}{VK} & AP & \textbf{0.063} & - & - & 0.006 & 0.021 \\
%						  & ROC &0.87 & - & - & 0.77 & \textbf{0.89}\\
\end{tabular}
\caption
{\label{tab:results}Results for each method on each graph. Results are significant (i.e. Variance is inferior to reported precision.}
\end{table}

To confirm this result, we plot the $precision@k$ score. In many applications, only the first few predictions are used, thus the relevance of this analysis. 

To perform this experiment, we select randomly pairs of nodes that are not linked in the training set until we obtain 1000 positive examples, i.e., edges that do appear according to the ground truth. which allows us to preserve a balance between positive and negative examples coherent with the dataset.

\begin{figure*}
    \centering
    \begin{subfigure}[h]{0.40\textwidth}
        \centering
        \includegraphics[width=\textwidth]{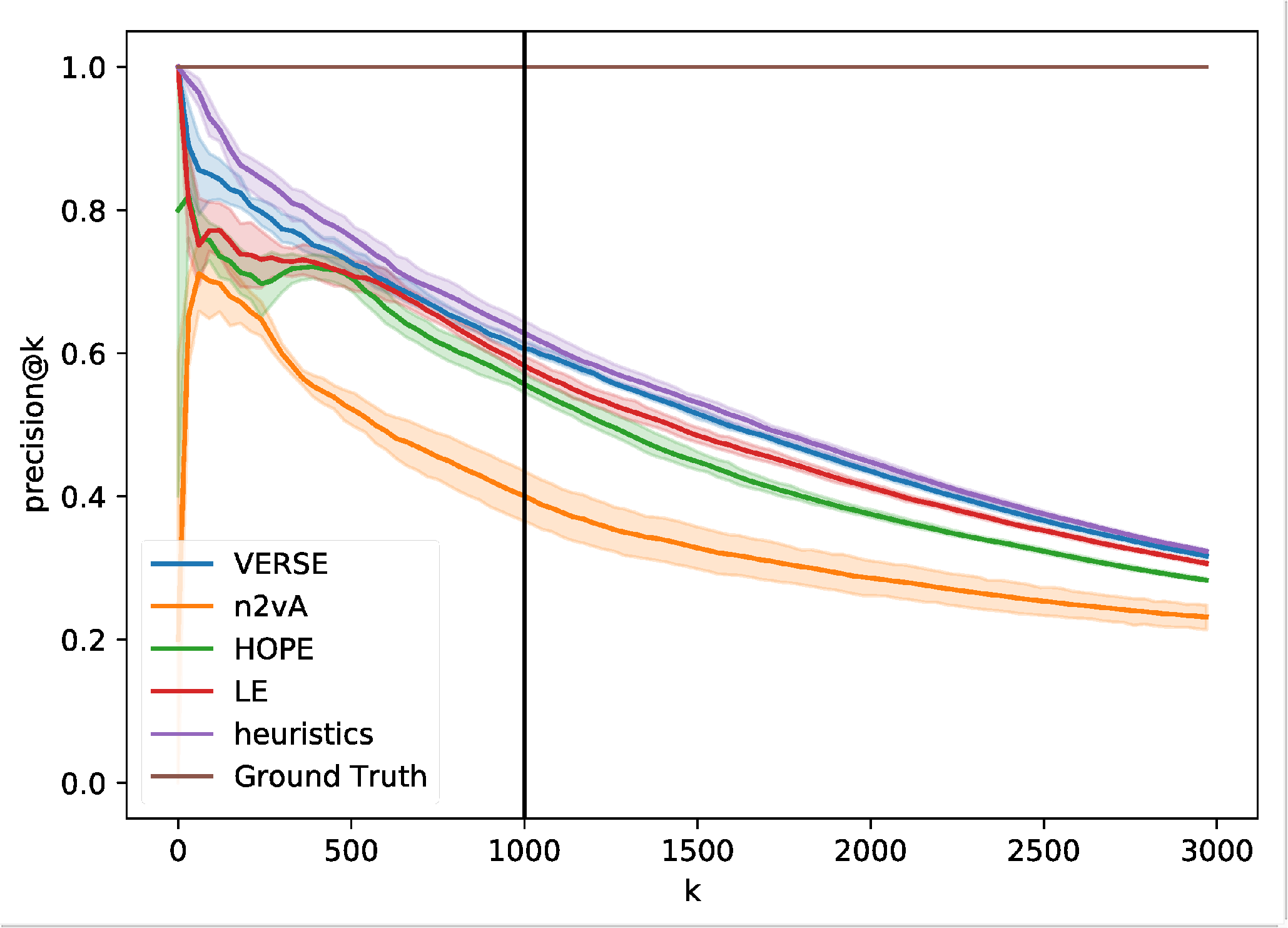}
        \caption{FACEBOOK}
    \end{subfigure}
    \begin{subfigure}[h]{0.40\textwidth}
        \centering
        \includegraphics[width=\textwidth]{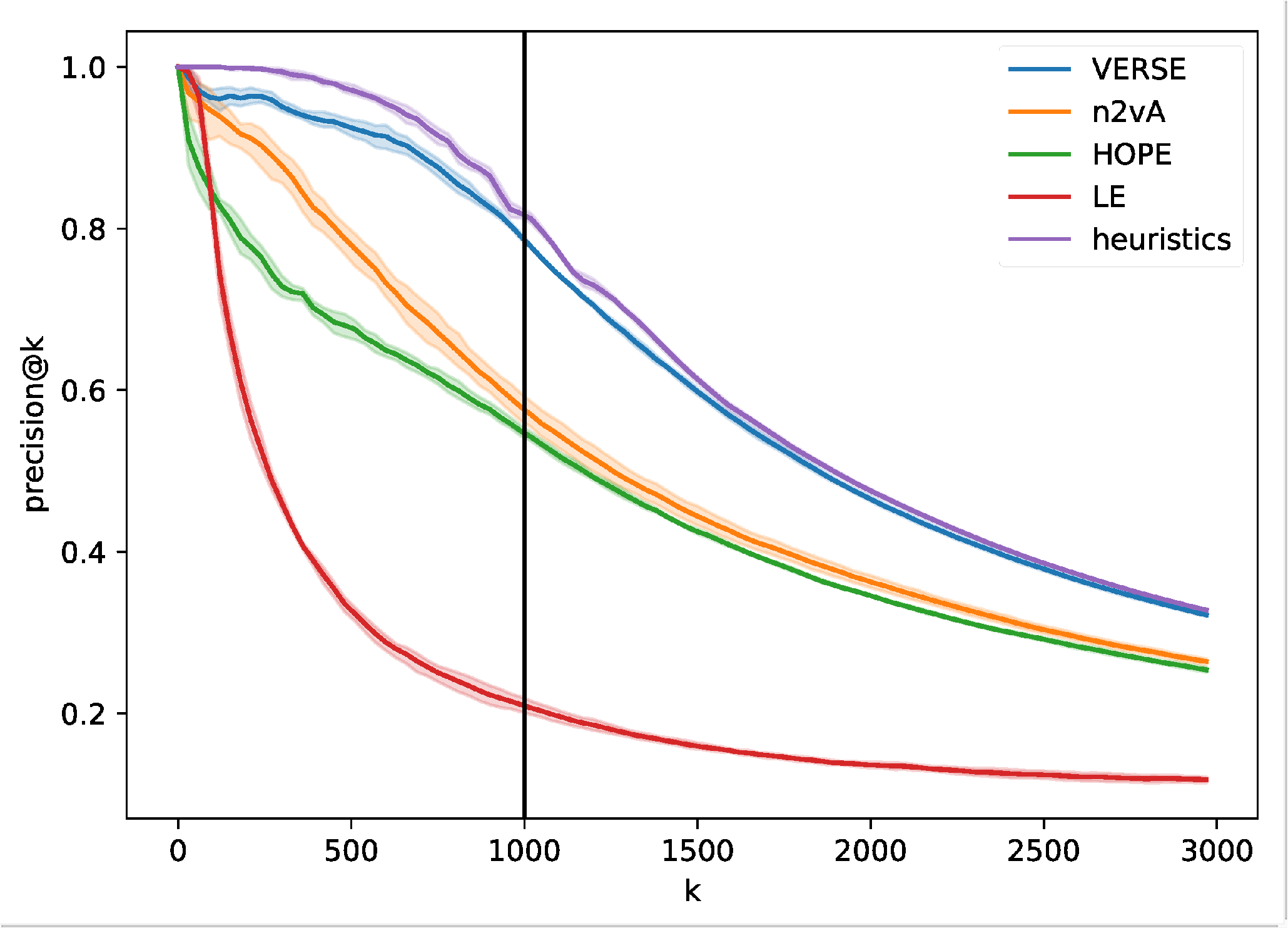}
        \caption{ASTROPH}
    \end{subfigure}
    
    \begin{subfigure}[h]{0.45\textwidth}
        \centering
        \includegraphics[width=\textwidth]{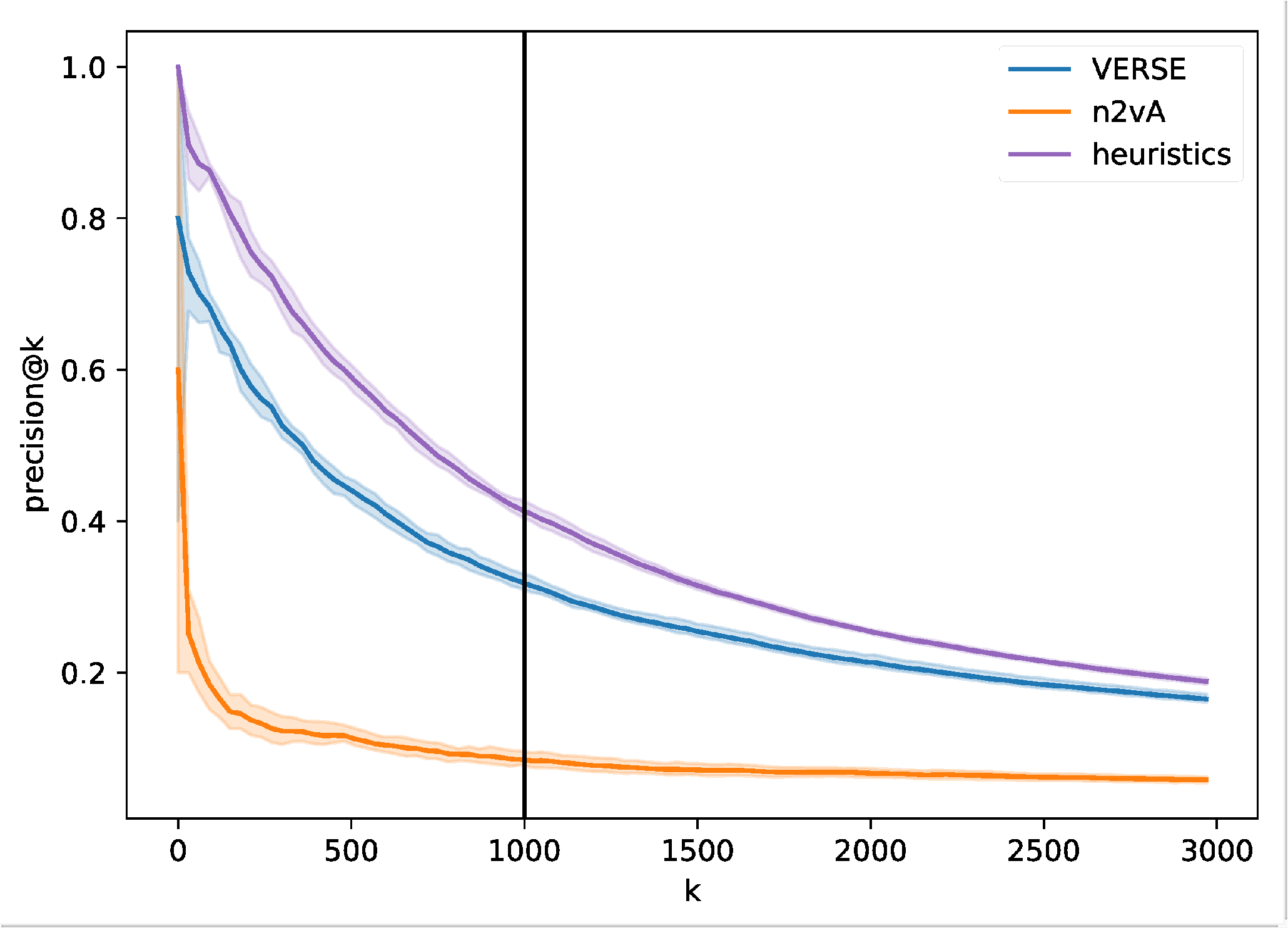}
        \caption{VK}
    \end{subfigure}
    \caption{Precision@k}

    \label{fig:precAtk}
\end{figure*}

Results are plotted in fig. \ref{fig:precAtk}. We observe that results are coherent with the scores obtained, i.e., heuristics and VERSE tend to give the best scores both for the first few predictions and for a realistic number of predictions (1000, corresponding to the number of real positive examples in the sample). In some settings, some algorithms do not yield useful predictions after the first few hundreds one (LE for ASTROPH dataset, node2vec for VK).

\section{Analysis of systematic biases}

New edges in a network can appear for different reasons, and between nodes with different properties. For instance, in a social network such as Twitter, a new follower relationship might be due to two close friends following each other, to a user following a raising celebrity they recently discovered, to strangers connecting because they think they have a common topic of interest, etc.

A link prediction algorithm might have a tendency to predict some types of links more than others, a phenomenon that we call \textit{systematic bias}. In this article, we focus on biases induced by the network topology. More particularly, we focus on three types of biases:
\begin{itemize}
	\item Graph distance
	\item Node degree
	\item Community structure
\end{itemize}

To evaluate the biases, we study the evolution of the fraction of new edges verifying a given property. Since predicted edges are ordered, from most probable to less probable, we define the $fraction@k$, corresponding to the ratio of pairs of nodes satisfying this property among the $k$ most likely edges. The dataset is selected as previously explained for the $precision@k$.

We define the reference value of $fraction@k$ as the value among all edges that do appear in the ground truth (positive examples in the test set). In the scenario of a perfect prediction, the $fraction@k$ curve should follow the ground truth scenario until 1000 (corresponding to the real number of observed edges in the test sample), and then move towards the value corresponding to the whole dataset (all pairs of nodes not linked in the training set).

\subsection{Graph distance}

\begin{figure*}
    \centering
    \begin{subfigure}[h]{0.4\textwidth}
        \centering
        \includegraphics[width=\textwidth]{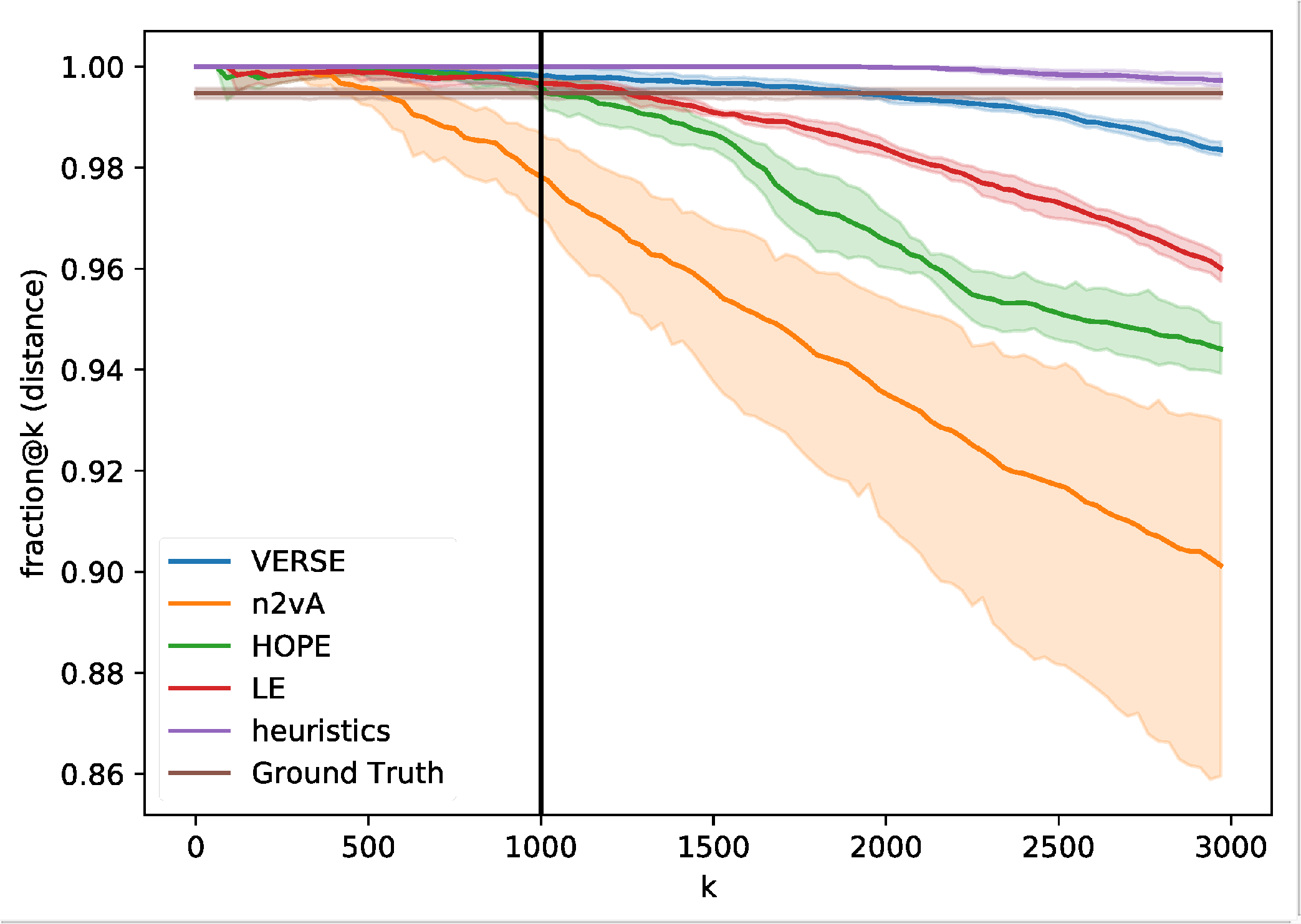}
        \caption{FACEBOOK}
    \end{subfigure}
    \begin{subfigure}[h]{0.4\textwidth}
        \centering
        \includegraphics[width=\textwidth]{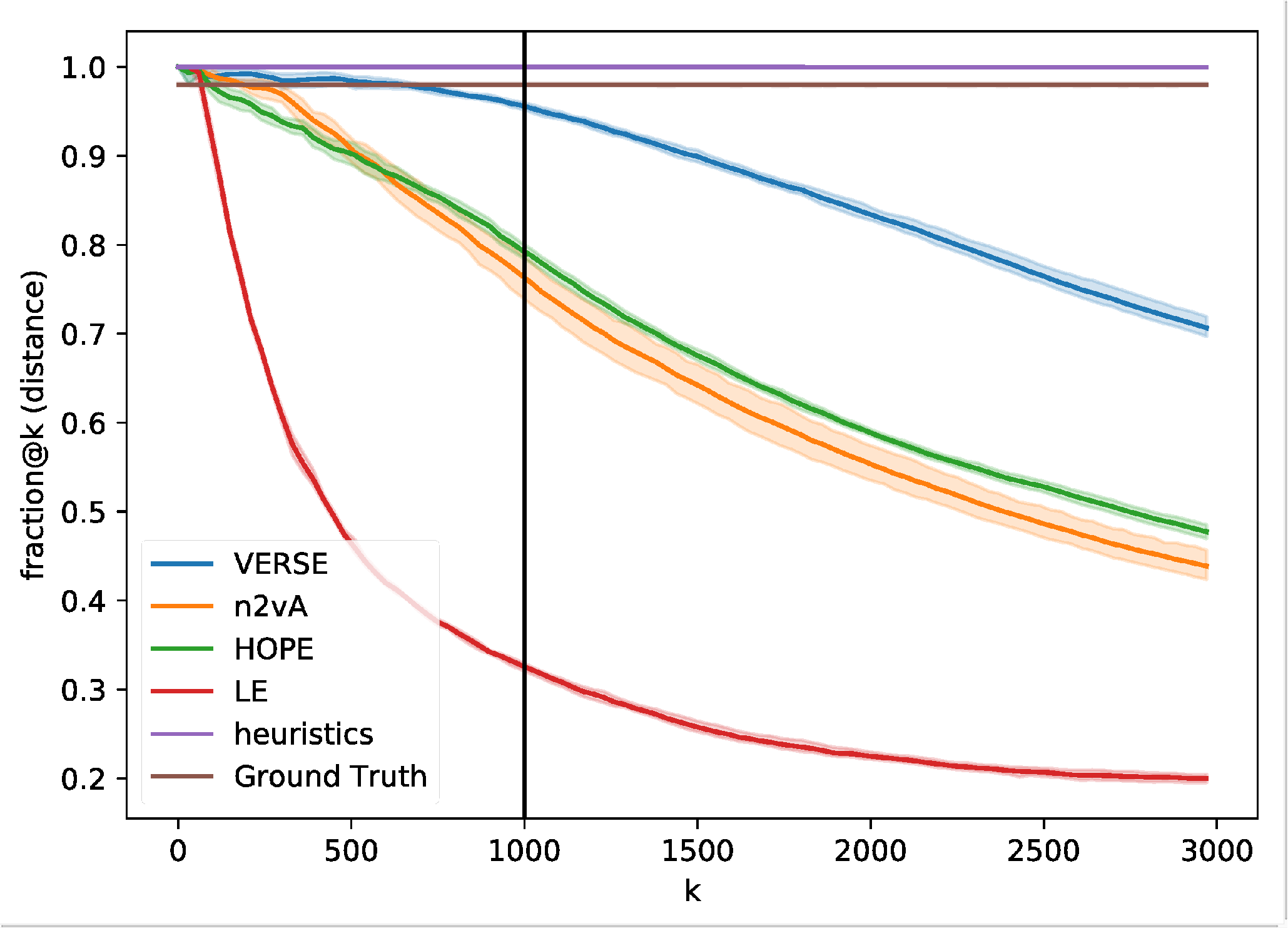}
        \caption{ASTROPH}
    \end{subfigure}
    \begin{subfigure}[h]{0.4\textwidth}
        \centering
        \includegraphics[width=\textwidth]{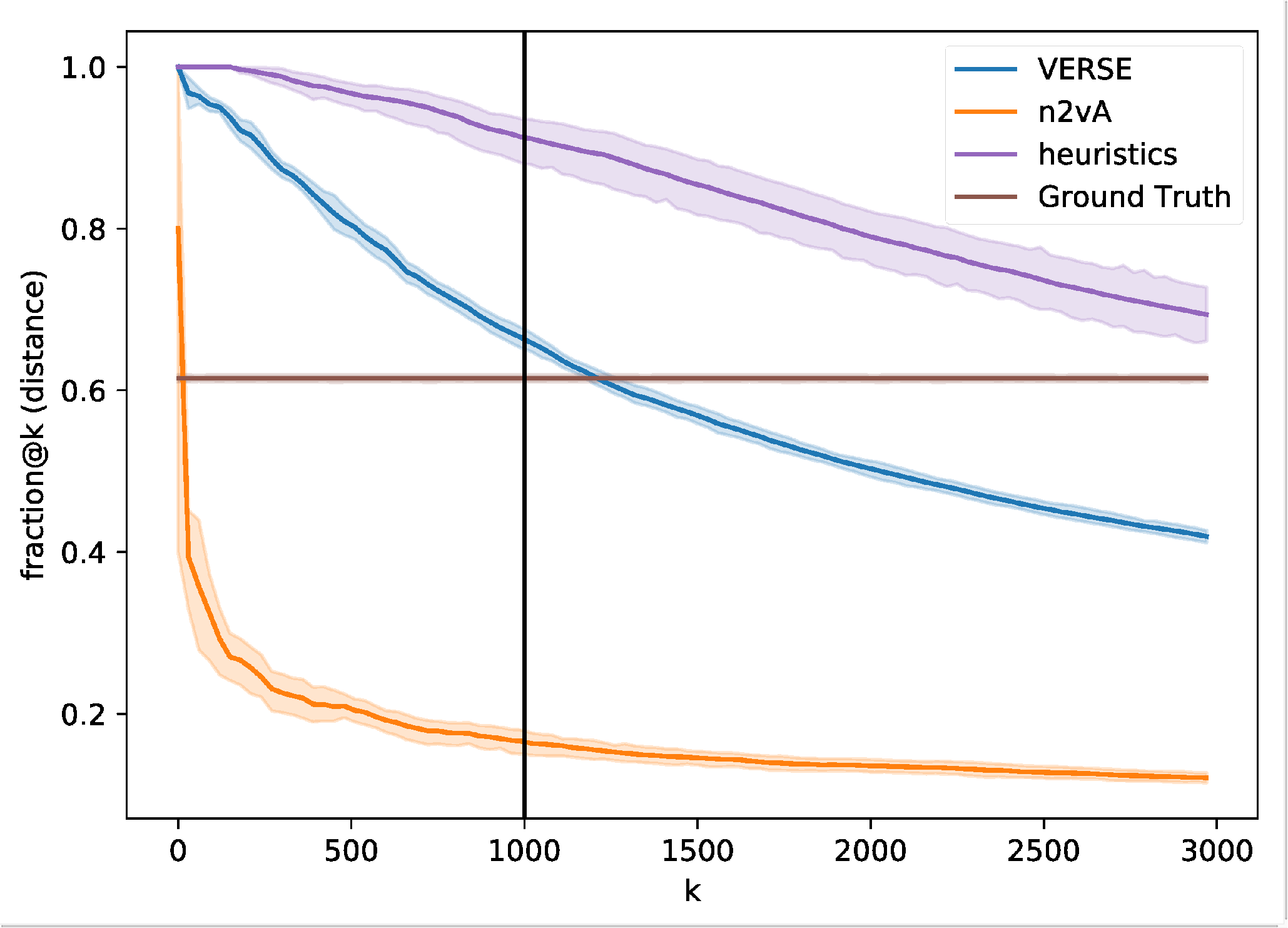}
        \caption{VK}
    \end{subfigure}
    \caption{Ratio@k for the graph distance. }

    \label{fig:distance}
\end{figure*}

Edges can appear between nodes that were \textit{close} or \textit{far} in the graph in term of graph distance, i.e. length of the shortest path between them. It has been observed that most edges appear between nodes at a short distance, a phenomenon often called \textit{triangle closure}. It is intuitively known in social network analysis by the saying "friends of my friends are my friends". Since the number of edges appearing at a distance more than two is usually very low, we consider only two cases: 
\begin{itemize}
	\item Short distance link: the new edge appear between nodes that were previously at distance two in the graph.
	\item Long distance link: the new edge appear between nodes that were previously at distance three or more.
\end{itemize}

Fig. \ref{fig:distance} presents the $fraction@k$ of short distance link for the different methods.

We can make the following observations:
\begin{itemize}
	\item In the ground truth, most edges appear between nodes at distance 2, although the value is much lower for VK dataset.
	\item The Heuristic-based approach is highly biased towards predicting short distance links.
	\item Most other approaches tend to be biased towards short distance links in the first (most probable) predictions, this value later decreases and the fraction often becomes lower than expected when the expected number of edges (1000) is reached.
\end{itemize}

\subsection{Node degree}

\begin{figure*}
    \centering
    \begin{subfigure}[h]{0.4\textwidth}
        \centering
        \includegraphics[width=\textwidth]{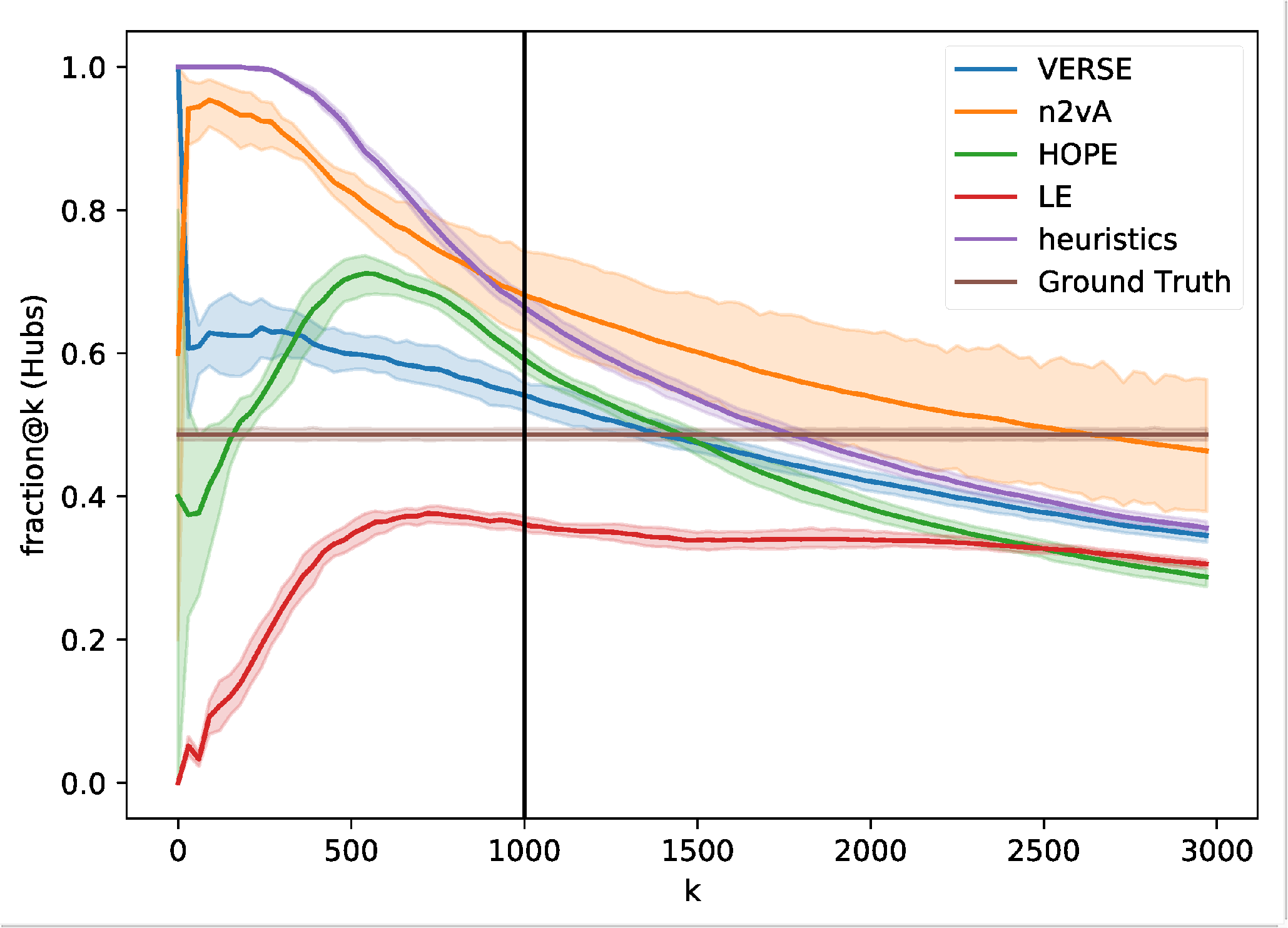}
        \caption{FACEBOOK}
    \end{subfigure}
    \begin{subfigure}[h]{0.4\textwidth}
        \centering
        \includegraphics[width=\textwidth]{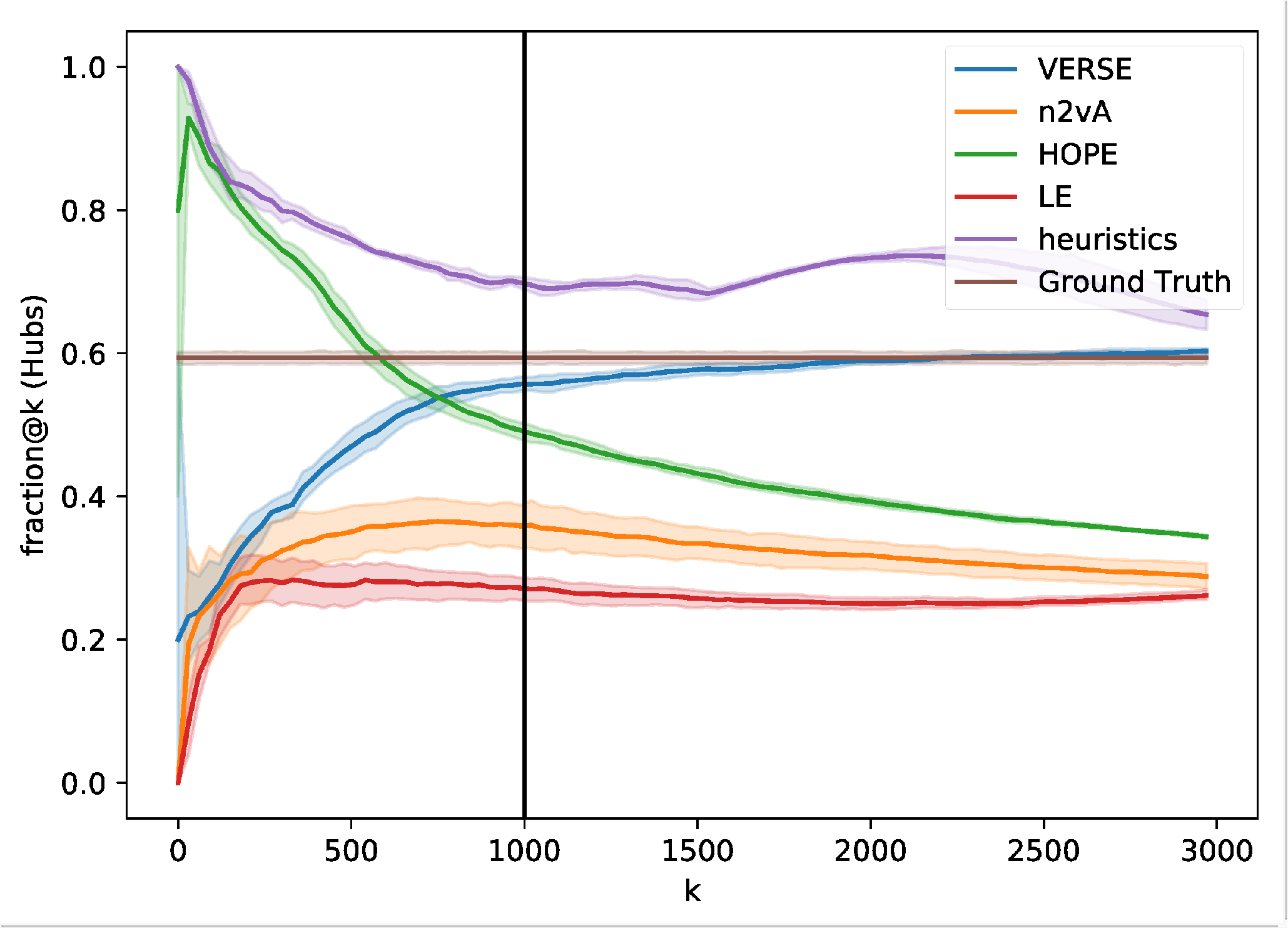}
        \caption{ASTROPH}
    \end{subfigure}
    
    \begin{subfigure}[h]{0.4\textwidth}
        \centering
        \includegraphics[width=\textwidth]{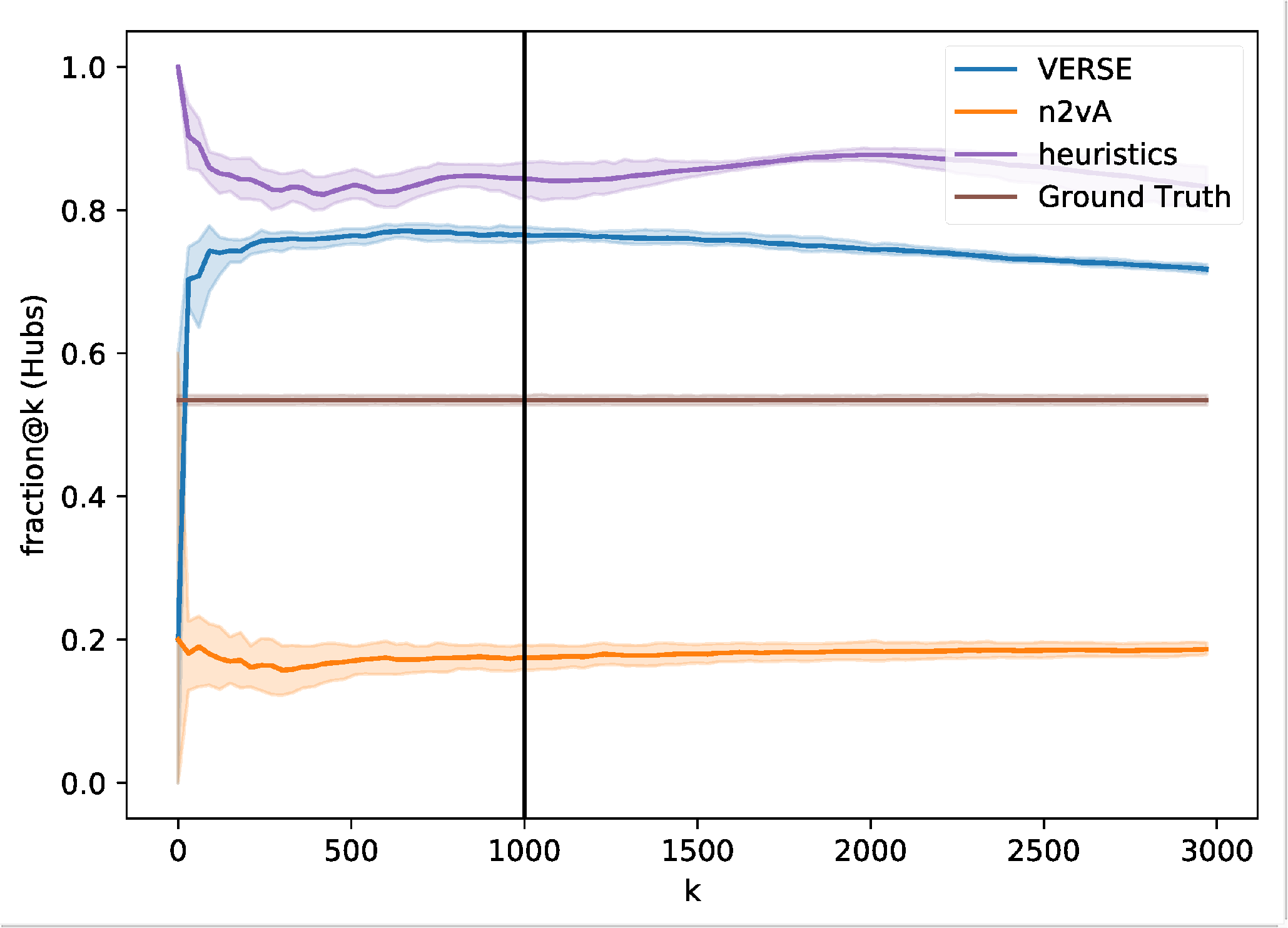}
        \caption{VK}
    \end{subfigure}
    \caption{Fraction@k for High degree nodes (hubs)}

    \label{fig:hubs}
\end{figure*}

Most real networks have heterogeneous degree distributions, that can often be approximated by scalefree distributions \cite{barabasi2009scale}. In those networks, there is a small fraction of nodes of high degrees that concentrate most of the edges. We define this class of nodes, called \textit{Hubs}, as the  10\% of nodes of highest degrees.

Fig. \ref{fig:hubs} presents the $fraction@k$ of new links that have at least a Hub among their endpoints.

We can make the following observations:
\begin{itemize}
	\item In the three studied datasets, the $fraction@k$ is between 0.5 and 0.6
	\item The Heuristic-based approach is highly biased towards high $fraction@k$, i.e. predicting too many edges involving hubs.
	\item One method seems to be clearly biased towards underestimation (LE), most other methods do not have clear tendencies.
	\item The VERSE method is the closest to the Ground Truth at the realistic threshold (1000) in all 3 settings.
\end{itemize}

\subsection{Community structure}

\begin{figure*}
    \centering
    \begin{subfigure}[h]{0.3\textwidth}
        \centering
        \includegraphics[width=\textwidth]{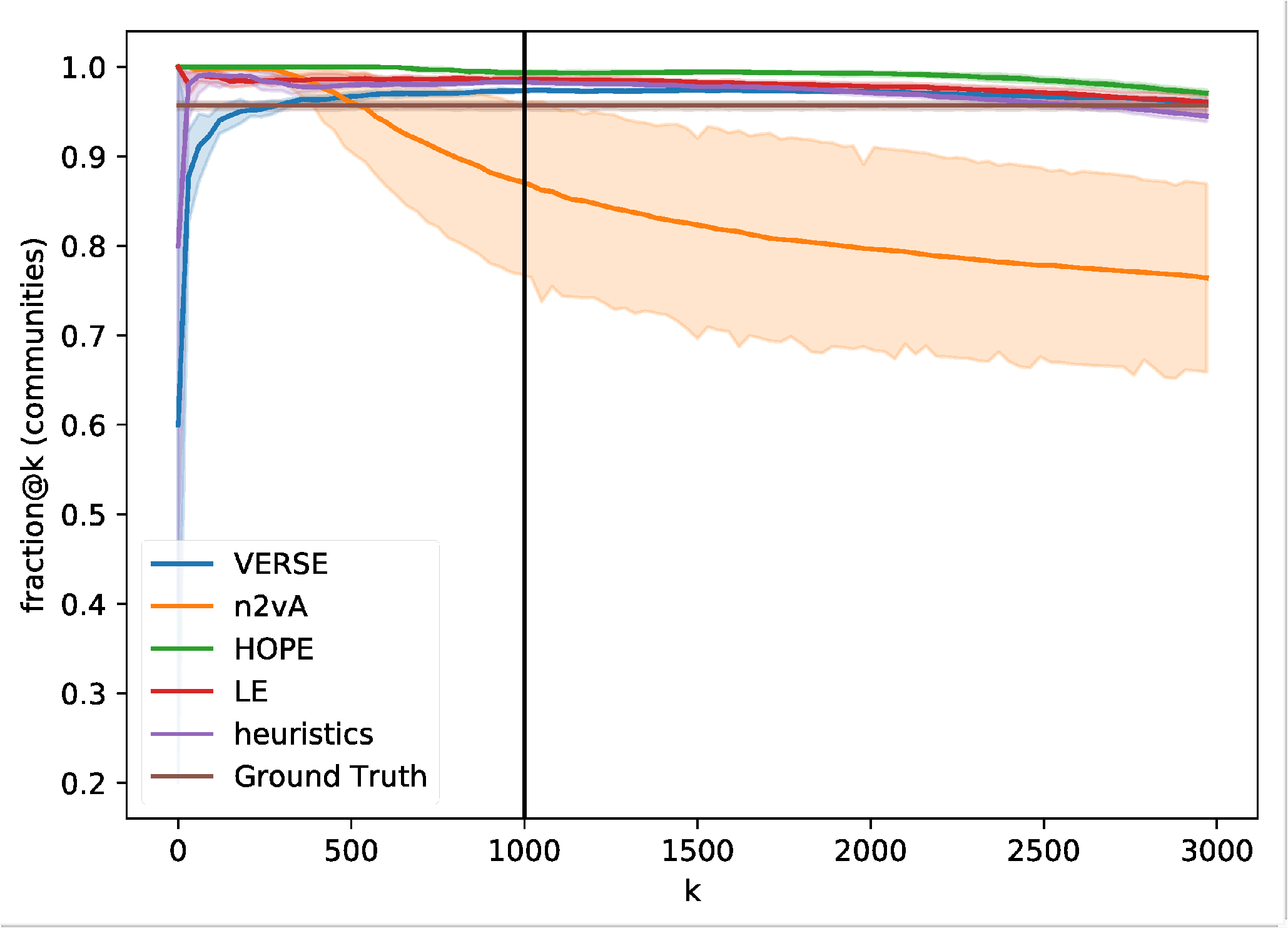}
        \caption{FACEBOOK}
    \end{subfigure}
    \begin{subfigure}[h]{0.3\textwidth}
        \centering
        \includegraphics[width=\textwidth]{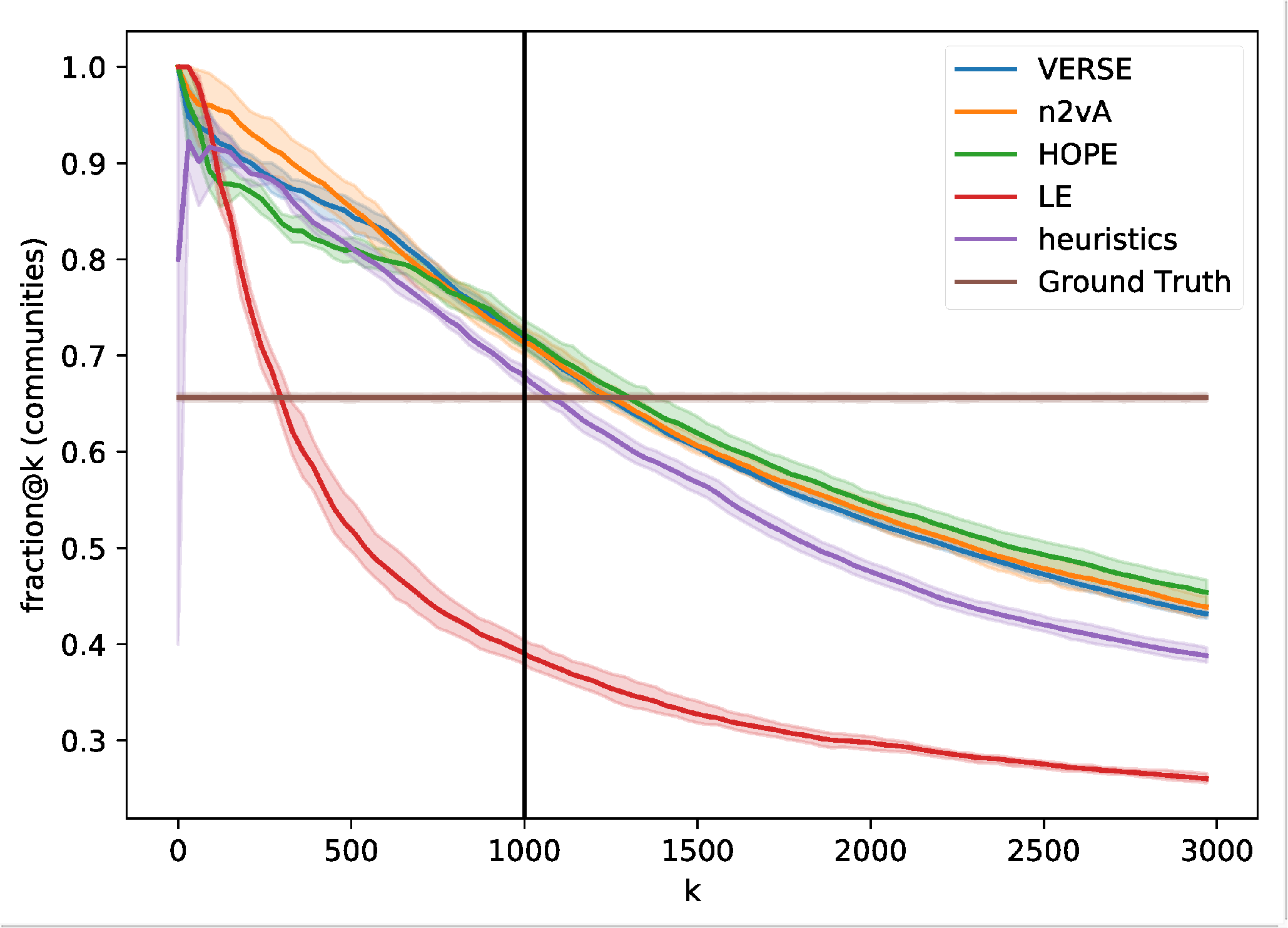}
        \caption{ASTROPH}
    \end{subfigure}
    \begin{subfigure}[h]{0.3\textwidth}
        \centering
        \includegraphics[width=\textwidth]{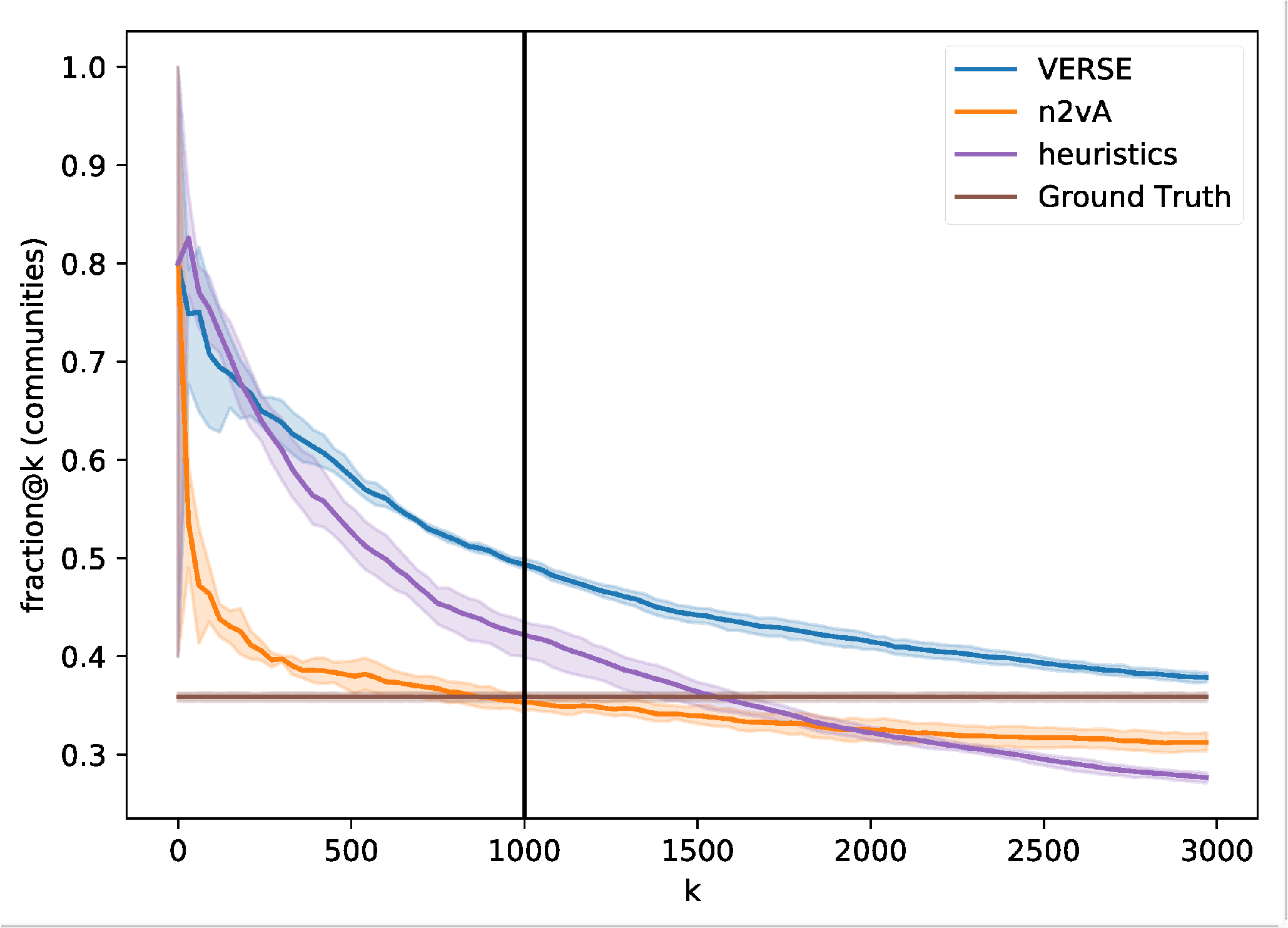}
        \caption{VK}
    \end{subfigure}
    \caption{Fraction@k for the community structure}

    \label{fig:communities}
\end{figure*}

Most real networks have mesoscale structures known as communities\cite{fortunato2010community}. The field of community detection and analysis is very active, and there is an intricate relationship between link prediction and community structure\cite{clauset2008hierarchical}. Because communities are dense groups of nodes, edges are more likely to appear inside communities than between them, a property that can be used both for link prediction and for community discovery in dynamic settings\cite{rossetti2018community}. 

We therefore distinguish between two types of edges, those appearing inside communities and those appearing between them.

To discover the community structure, we apply the Louvain algorithm \cite{blondel2008fast}, which is the most often used in the literature. This algorithm is based on Modularity optimization, thus, by definition, discover communities denser than the rest of the network. Note that the problem could be studied in more details using overlapping communities, that can introduce different effects, such as higher probabilities of edges on the overlapping parts \cite{yang2012community}

Fig. \ref{fig:communities} presents the $fraction@k$ of new links that appear inside communities, as found by the Louvain Algorithm.

We can make the following observations:
\begin{itemize}
	\item In different datasets, the fraction of edges in the ground truth appearing inside communities varies widely.
	\item The Heuristic-based approach is systematically biased towards high $fraction@k$, i.e. predicting too many edges inside communities.
	\item Most methods (except n2v in FACEBOOK and LE in ASTROPH) are also biased towards overprediction of internal edges. 
	\item The bias is particularly strong among the first few prediction
\end{itemize}

\section{Discussion: effect of biases on recommender systems}
In recent years, both the scientific community and the civil society have started to pay attention to the problem of biased results provided by machine learning algorithms in real-life applications. Two problems have particularly been identified:
\begin{itemize}
	\item The filter bubble phenomenon is known to occur at least in two settings: in Online Social Networks such as Facebook and Twitter, and on web search engines such as Google or Bing. The problem comes from a reinforcement phenomenon: machine learning algorithms learn the preferences of users, and show them more results according to these preferences. As a consequence, users see results that are less diverse than they should, but rather biased towards the opinions or interests that the algorithm has inferred they had in the beginning. This process has been accused to amplify the \textit{polarization} phenomenon in political and social opinion \cite{pariser2011filter}.
	\item The problem of fairness of algorithms \cite{berk2017fairness} arrises in many real-life applications of machine learning algorithms. In its most famous occurrences, a decision is taken by an algorithm that will impact individuals (who gets a loan, a job, is accepted at a university, is put in a list of potential terrorists, etc.), but the decisions taken by the algorithms are unfair towards a category of people. Typically, the algorithm learns that an ethnic or social group --or, if not present in the data, another highly correlated attribute such as locations or name-- correlates with unfavorable outcomes, and therefore learns to discriminate based on this property.
\end{itemize}

These two problems are now widely known, and solutions have been proposed to mitigate them\cite{bolukbasi2016man}, in particular by ensuring the preservation of some identified properties. 

The systematic biases highlighted in the present article however have several interesting features:
\begin{itemize}
	\item They arise without labeled data, but simply due to the network structure
	\item The problem is not that unwanted problematic correlations are preserved, but rather that some network properties that were not explicitly required to be preserved are lost.
\end{itemize}

We can note that the efficiency of the link prediction is not necessarily linked to the absence of bias, on the contrary, methods based on heuristics clearly yield the best results in term of overall link prediction compared with recent techniques based on embeddings. However, in term of biases, we have observed that heuristics usually suffer from the highest biases. Furthermore, those biases are constantly of the same types: they tend to favor nodes at a short distance in the graph, large nodes, and nodes belonging to the same communities. In other terms, these biases can be associated with a \textbf{loss in diversity}. In a social network, recommendations of new contacts will favor exaggeratedly either the most similar profiles (same community, closer distance) or the most popular profiles (hubs), but will fail to identify more nuanced possibilities that would add diversity. In product or content recommendation, an algorithm based on link prediction with heuristics will propose the items the most similar to those already consumed, or the most consumed overall, ignoring the relevant possibilities in-between.

New methods based on embeddings might be part of the solution to this problem, as some of them tend to reduce those biases. In particular, it seems that the recent VERSE algorithm, that offers the best overall results among embedding techniques for link prediction accord to scores and $precision@k$, also consistently show lesser biases than previous techniques.

On top of that, one could follow techniques already proposed\cite{bolukbasi2016man} to fight against the fairness of algorithm issue, for instance by training separately for the different properties to preserve (e.g., training different classifiers to predict edges at distance 2 and at distance 3 or more).

\section{Conclusion}
We have shown that the most widely used techniques for link prediction suffer from systematic biases towards some network properties, and that those biases might play a role in the filter bubbles and algorithmic fairness problems. 

We have also identified that some of the most recent techniques proposed based on graph embedding are able to somewhat reduce that problem, although they do not yet manage to improve over heuristic based methods for the overall quality of the link prediction.

We think that this work can be extended in two directions: 
\begin{itemize}
	\item On the one hand, we need to understand where do those biases come from. While it can be intuitively understood for heuristics --the features are not learned but selected according to prior knowledge, and they have been designed with triangle closure and preferential attachment properties in mind-- the reason why feature learning based methods have similar biases is, to the best of our knowledge, unknown. 
	\item On the other hand, we must develop link prediction methods adapted to mitigate or eliminate those systematic biases. Ideally, such a method should not only remove the three biased identified in this article, but ensure using statistical methods that the network formed by adding the predicted edges has a similar network structure profile than the original network.
\end{itemize}

\bibliographystyle{spmpsci}
\bibliography{sample}
%\begin{thebibliography}{6}
%%
%
%\bibitem {smit:wat}
%Smith, T.F., Waterman, M.S.: Identification of common molecular subsequences.
%J. Mol. Biol. 147, 195?197 (1981). \url{doi:10.1016/0022-2836(81)90087-5}
%
%\bibitem {may:ehr:stein}
%May, P., Ehrlich, H.-C., Steinke, T.: ZIB structure prediction pipeline:
%composing a complex biological workflow through web services.
%In: Nagel, W.E., Walter, W.V., Lehner, W. (eds.) Euro-Par 2006.
%LNCS, vol. 4128, pp. 1148?1158. Springer, Heidelberg (2006).
%\url{doi:10.1007/11823285_121}
%
%\bibitem {fost:kes}
%Foster, I., Kesselman, C.: The Grid: Blueprint for a New Computing Infrastructure.
%Morgan Kaufmann, San Francisco (1999)
%
%\bibitem {czaj:fitz}
%Czajkowski, K., Fitzgerald, S., Foster, I., Kesselman, C.: Grid information services
%for distributed resource sharing. In: 10th IEEE International Symposium
%on High Performance Distributed Computing, pp. 181?184. IEEE Press, New York (2001).
%\url{doi: 10.1109/HPDC.2001.945188}
%
%\bibitem {fo:kes:nic:tue}
%Foster, I., Kesselman, C., Nick, J., Tuecke, S.: The physiology of the grid: an open grid services architecture for distributed systems integration. Technical report, Global Grid
%Forum (2002)
%
%\bibitem {onlyurl}
%National Center for Biotechnology Information. \url{http://www.ncbi.nlm.nih.gov}
%
%
%\end{thebibliography}
\end{document}